\documentclass[superscriptaddress, reprint,preprintnumbers,nofootinbib, amsmath,amssymb,aps]{revtex4-2}
\usepackage[utf8]{inputenc}
\usepackage{graphicx}% Include figure files
\usepackage{dcolumn}% Align table columns on decimal point
\usepackage{bm}% bold math
\usepackage{amssymb,bm,mathrsfs,bbm,amscd}
\usepackage{subfig,color,slashed}
\usepackage{amssymb}
\usepackage{multirow}
\usepackage{cancel}
\usepackage{caption}
\usepackage{changes}
\usepackage{hyperref}
\usepackage{booktabs}
\usepackage{textcase}
\usepackage{url}
\usepackage[normalem]{ulem}
\usepackage{mathtools}


\definechangesauthor[name=Dong Woo Kang, color=magenta]{DW}

\def\deelema{\textsf{DeeLeMa}~}

\usepackage{hhline}% http://ctan.org/pkg/hhline
\usepackage{soul}

\newcommand*{\bigE}{\scalebox{1.2}{\ensuremath{\mathcal{E}}}}
\newcommand*{\bigB}{\scalebox{1.2}{\ensuremath{\mathcal{B}}}}

\begin{document} 
\preprint{KIAS-P22085}
\preprint{CERN-TH-2022-218}

\title{\texorpdfstring{\deelema:}{deelema} Missing information search with Deep Learning for Mass estimation}
 
\author{Kayoung Ban}
%\email{ban94gy@yonsei.ac.kr}
\affiliation{Department of Physics and IPAP, Yonsei University, Seoul 03722, Republic of Korea}

\author{Dong Woo Kang}
%\email{dongwookang@kias.re.kr}
\affiliation{Korea Institute for Advanced Study, Seoul 02455, Republic of Korea}
\affiliation{Theoretical Physics Department, CERN, Geneva, Switzerland}

\author{Tae-Geun Kim}
%\email{tg.kim@yonsei.ac.kr}
\affiliation{Department of Physics and IPAP, Yonsei University, Seoul 03722, Republic of Korea}
\author{Seong Chan Park}
%\email{sc.park@yonsei.ac.kr}
\affiliation{Department of Physics and IPAP, Yonsei University, Seoul 03722, Republic of Korea}
\author{Yeji Park}
%\email{yeji.park@yonsei.ac.k}
\affiliation{Department of Physics and IPAP, Yonsei University, Seoul 03722, Republic of Korea}

\begin{abstract}

We introduce \textsf{DeeLeMa}, a deep learning-based network for the analysis of energy and momentum in high-energy particle collisions. This novel approach is specifically designed to address the challenge of analyzing collision events with multiple invisible particles, which are prevalent in many high-energy physics experiments. \textsf{DeeLeMa} is constructed based on the kinematic constraints and symmetry of the event topologies.  We show that \textsf{DeeLeMa} can robustly estimate mass distribution even in the presence of combinatorial uncertainties and detector smearing effects. The approach is flexible and can be applied to various event topologies by leveraging the relevant kinematic symmetries. This work opens up exciting opportunities for the analysis of high-energy particle collision data, and we believe that \textsf{DeeLeMa} has the potential to become a valuable tool for the high-energy physics community. 

\end{abstract}

\maketitle
\flushbottom

\section{Introduction \label{sec:intro}}

Despite the numerous neutrinos generated during particle collisions, the detectors at the Large Hadron Collider (LHC) are unable to observe them directly~\cite{ATLAS:2023gsl, CMS:2023qyl}. In addition to neutrinos, other elusive particles such as Dark Matter candidates, including Weakly Interacting Massive Particle (WIMP) \cite{Jungman:1995df, Bertone:2004pz}, Axions~\cite{Kim:1979if, Dine:1981za}, are also challenging to detect as they pass through the detector without leaving discernible signals \cite{CMS:2021juv,Gonski:2022gnz}. Such entities are termed as `invisible particles' in the realm of collider physics. Their existence isn't directly observed but is inferred by leveraging the principles of energy and momentum conservation, which highlight discrepancies in momentum or energy within an event.

The LHC, like other hadronic collider experiments, measures the scattering processes involving the partonic constituents of hadrons; Within this context, the reconstruction of the longitudinal component of missing momentum along the beam axis (referred to as the longitudinal direction) poses a substantial challenge. Furthermore, the formidable nature of this endeavor becomes particularly pronounced when multiple invisible particles are simultaneously generated within the same event. This challenging issue is conventionally called the ``{\it missing information problem}" of invisible particles.

Researchers commonly employ `transverse' quantities to address the challenge from the longitudinal information. These transverse quantities are defined along directions perpendicular to the beam axis and include the transverse momentum ($p_T =\sqrt{{\vec{p}_\perp}^2}$) and transverse energy ($E_T \equiv \sqrt{m^2-p_T^2}$) as observable parameters. Over the past decade or more, many kinematic variables have been devised and proposed, primarily tailored for the experiments at the LHC, such as the stransverse mass or the Cambridge $M_{T2}$~\cite{Barr:2003rg, Lester:1999tx, Cho:2008cu}, $M_2$~\cite{Barr:2011xt, Mahbubani:2012kx, Cho:2014naa}, and their extensions~\cite{Burns:2008va, Barr:2009, Konar:2009qr, Konar:2009wn, Cho:2008cu}.  However, it is worth noting that introducing more complex kinematic variables while aiding in obtaining missing information can also introduce additional complexities in data analysis. The precision of these variables may not always meet the desired level due to inherent complexities and uncertainties, including combinatorial errors and detector effects.
For a comprehensive overview, see e.g., Ref.~\cite{Franceschini:2022vck}.

This paper introduces an innovative approach to address the challenges posed by missing information problems in collider physics.~\cite{Nojiri:2003tu, Kawagoe:2004rz, Cheng:2007xv, Cheng:2008mg, Cheng:2009fw, Webber:2009vm, Kim:2019prx} Instead of relying on intricate kinematic variables, our proposed method leverages the power of Deep Neural Networks (DNNs), capitalizing on the recent rapid advancements in machine learning techniques~\cite{Feickert:2021ajf, Radovic:2018dip, Shanahan:2022ifi, Karagiorgi:2021ngt, Dong:2022trn, Kim:2021pcz, Haq:2022oir, Alves:2022gnw, Lim:2020igi, Chakraborty:2020yfc}.

DNNs have emerged as a versatile tool capable of handling vast datasets and capturing intricate correlations among diverse features. This capability renders them exceptionally well-suited to tackle the complexities associated with missing information. Our newly developed kinematics-solving machine integrates the physical conditions and symmetries inherent in event shapes, is named ``\deelema." This acronym, derived from ``Deep Learning for Mass Estimation," encapsulates the essence of our machine's function. \deelema represents a cutting-edge approach to the problem of kinematics estimation in collider physics, promising more robust and accurate results compared to traditional methods reliant on complex kinematic variables.
The detail of the architecture is presented in the GitHub page\footnote{\url{https://github.com/Yonsei-HEP-COSMO/DeeLeMa}\label{github}}, where one can download \deelema code with examples.

%%%%%%%%%%%%%%%%%%%%%%%%%%%%%%%%%%%%%%%%%%%%%%%%%%%%%%%%%%%%%%%%%%%%%%%%%%%%%%%%%%%%%%%%%%%%%%%%%%%%%%%%%%%%%%%%%%%%%%%%%%%%%%%%%%%%%%%%%%
% \section{\texorpdfstring{\textsf{D\MakeLowercase{ee}L\MakeLowercase{e}M\MakeLowercase{a}}}{DeeLeMa} Modeling}
\section{\texorpdfstring{\textsf{D\MakeLowercase{ee}L\MakeLowercase{e}M\MakeLowercase{a}}}{DeeLeMa} Framework}

  \begin{figure}[t]
	\includegraphics[width=0.49\textwidth]{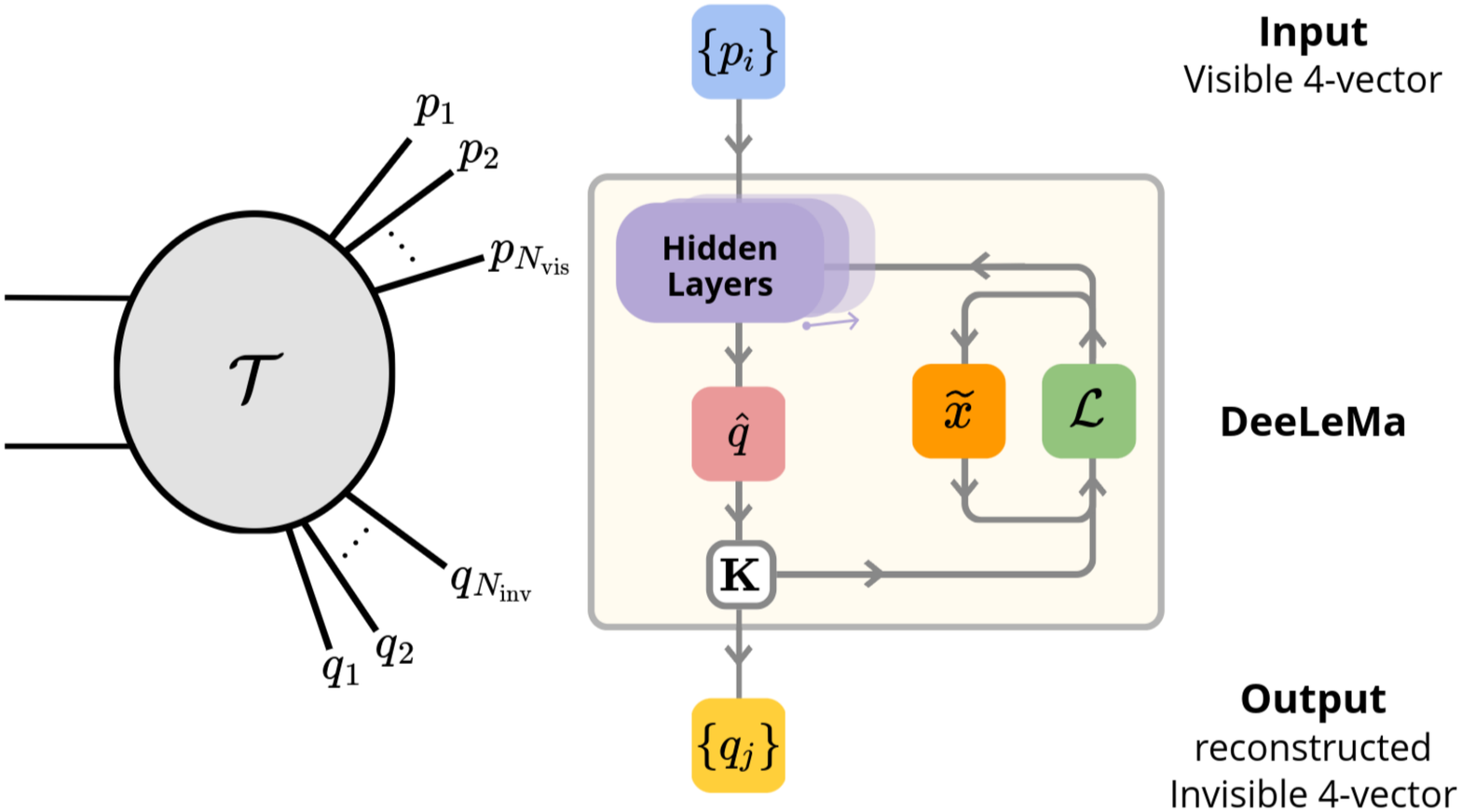}
	\caption{\small General event shape with visible ($\{p_i\}_{i=1,2,\cdots, N_{\rm vis}}$) and invisible ($\{q_j\}_{j=1,2,\cdots, N_{\rm inv}}$) momenta in the final state. 
	}
	\label{model1}
\end{figure}

Our study is dedicated to unveiling concealed information within the complex landscape of high-energy collider events. We aim to achieve this objective by harnessing observable data, specifically the four-momenta of detected particles. Event topology, symmetry principles, and the steadfast application of conservation laws furnish constraints on the kinematic variables governing these events.

To illustrate, we examine a cascade event configuration consisting of $N_{vis}$ visible particles and $N_{inv}$ invisible particles in the final state, which can be succinctly represented as:
$$ pp \to \overbrace{v_1 v_2 \cdots v_{N_{vis}}}^{N_{vis}} \overbrace{i_1 i_2 \cdots i_{N_{inv}}}^{N_{inv}}.$$
Our primary goal is to utilize the input information encapsulated in the four-momenta of visible particles, denoted as ${p_i}$ for $i=1,2,\cdots, N_{vis}$, to precisely determine the momenta of each invisible particle in the final state, which we designate as ${q_j}$ for $j=1,2,\cdots, N_{inv}$.
Nonetheless, it is crucial to notice that this kinematic problem becomes mathematically underdetermined when the count of unknown variables, $N_{inv}$, surpasses the constraining relationships governing each event's momenta.

Utilizing a physics-informed machine learning approach, we build a model that decodes concealed information in collider events under a given event topology. Central to this approach are two functions: $\mathcal{L}$, our loss function for neural optimization, and the function \textbf{K} serves as a mechanism that encapsulates kinematic relationships crucial for reconstructing the momenta of invisible particles. 
These functions are based on physical relations such as the on-shell mass conditions for the intermediate particles and the constraints on the transverse momentum.
The structure of our DNN machine is schematically depicted in Fig.~\ref{model1}:
\begin{itemize}

\item The event topology of the specific event is $\mathcal{T}$, and the kinematic relations among momenta are encapsulated in \textbf{K}.  

\item The input for \deelema is the visible information from the measured momenta $\{p_i\}$, $i=1,2,\cdots, N_{vis}$. 

\item The expected output from \deelema  is the reconstructed momenta of the invisible particles $\{q_j\}$, $j=1,2,\cdots, N_{inv}$.

\item The loss function $\mathcal{L}$ enforces the machine to learn to reconstruct the invisible information under the given event topology $\mathcal{T}$ and the kinematic relations \textbf{K}.

\end{itemize}

Additionally, we introduce the auxiliary parameters $\tilde{x}$ which act to force target physical variables $x$ (i.e. invariant mass) to converge into a single value for all training events. The corresponding auxiliary parameters $\widetilde{x}$ appear globally in all events, allowing the neural network to learn that the events come from the same physical process. 
Thus, they are introduced as global, trainable parameters based on prior knowledge from $\mathcal{T}$. 
Consequently, \deelema works to optimize the reconstruction of invisible momenta by minimizing the loss function $\mathcal{L}$, which is defined in terms of the reconstructed kinematic quantities $\hat{q}$ and the auxiliary parameters $\widetilde{x}$, subject to the kinematic relations \textbf{K}.

\section{\texorpdfstring{\textsf{D\MakeLowercase{ee}L\MakeLowercase{e}M\MakeLowercase{a}}}{DeeLeMa} for Pair Production Process }

In this section, our primary focus lies on the pair production of mother particles during particle collisions, where each of these particles subsequently decays, following identical decay chains. Under such circumstances, the scenario involves an even number of both visible and invisible particles, denoted as $(N_{vis}, N_{inv}) = (2n,2m)$. Here, the terms $n$ and $m$ correspond to the visible and invisible particles in each respective branch. Exploiting the inherent symmetry of this situation, we find that there are precisely $8m$ unknown components originating from the $2m$ invisible four-momenta, along with $(n+m+2)$ constraints stemming from kinematic relations.

Mathematically speaking, the system becomes solvable when the condition $(n+m+2) \geq 8m$ or equivalently $n\geq 7m-2$ is satisfied. A pertinent illustration is the case of $m=1$, wherein a single invisible particle emerges in each of the decay chain branches. In this scenario, the system can be effectively solved when $n\geq 5$. It is noteworthy to mention that earlier analyses on systems involving $n=3, m=1$ have been documented in previous works (see \cite{Cheng:2008mg, Cheng:2009fw, Webber:2009vm}), particularly when multiple events of the identical process were considered.

%%%%%%%%%%%%%%%%%%%%%%%%%%%%%
\begin{figure}[h]
\includegraphics[width=0.35\textwidth]{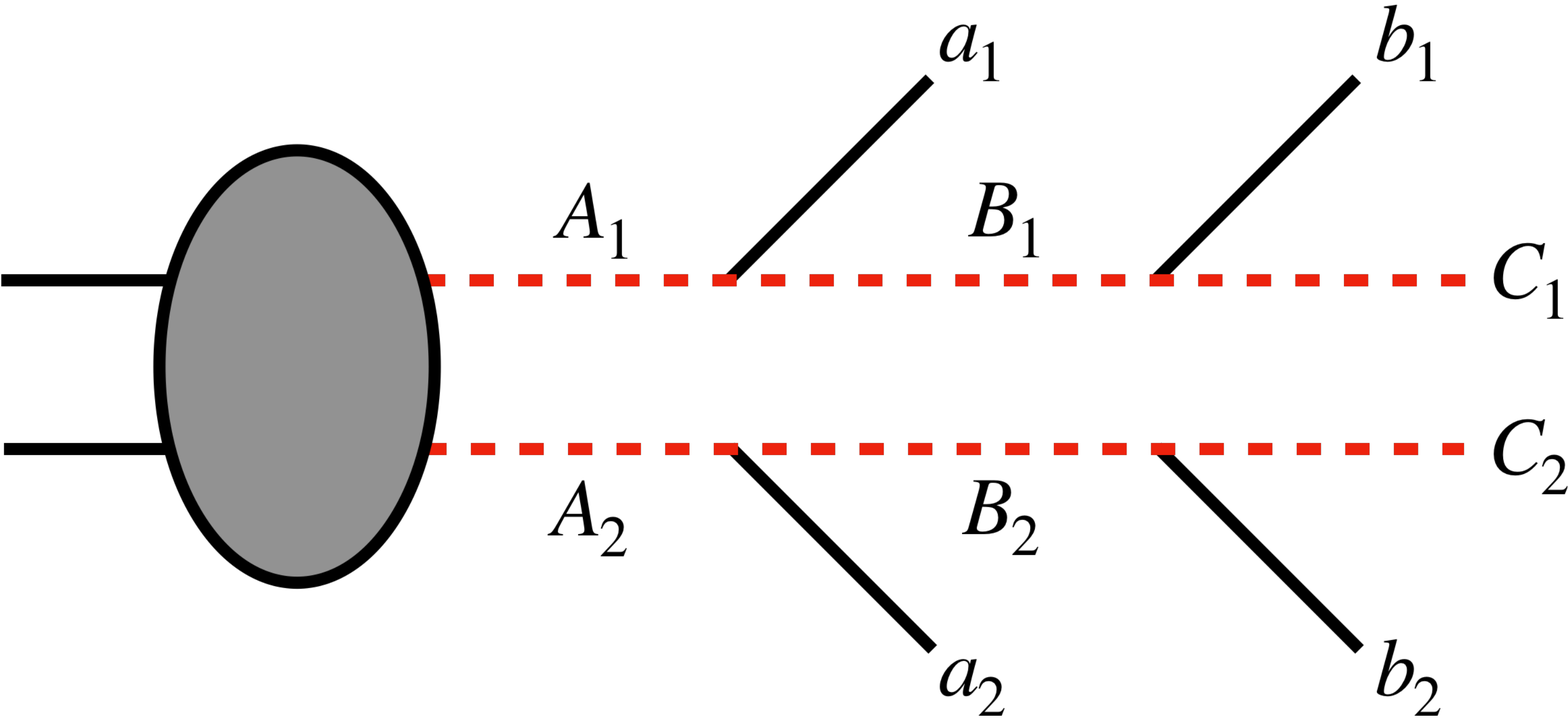}
 \caption{\label{topology}\small Symmetric event topology for $N_{vis}=4, N_{inv}=2$.}
 \end{figure}
%%%%%%%%%%%%%%%%%%%%%%%%%%%%%

We now delve into a challenging `unsolvable' problem characterized by the parameters $n=2$ and $m=1$, visually represented in Fig.~\ref{topology}. This specific configuration corresponds to an event topology of $(N_{vis}, N_{inv}) =(4,2)$. A prominent example of this event topology is found in the dilepton process of $t\bar{t}$ events, where both top quarks undergo leptonic decay, leading to $t\to bW \to b (\ell \nu_\ell)$. In a more general context, we contemplate the pair production of mother particles, denoted as $A_1$ and $A_2$, with subscripts 1 and 2 signifying the respective branches of decay. Each $A_i$ subsequently decays into a visible particle $a_i$ and an intermediate heavy state $B_i$. Ultimately, $B_i$ undergoes a semi-invisible decay into a visible particle $b_i$ and an invisible particle $C_i$ in branches $i=1$ and $i=2$.
The event can be succinctly expressed as:
\begin{align*}
pp \to A_1 A_2 \to (a_1 (p_{a_1}) B_1) (a_2 (p_{a_2}) B_2), \\
B_1 \to b_1 (p_{b_1}) C_1 (q_1),\\
B_2 \to b_2 (p_{b_2}) C_2 (q_2).
\end{align*}

Here, $p_{a_i}$ and $p_{b_i}$ symbolize the momenta of visible particles, while $q_i$ represents the momentum of the corresponding invisible particles $C_i$. Despite the apparent simplicity of this event topology, it is fundamentally underdetermined from a kinematic perspective, rendering the separate measurement of each invisible particle's momentum unattainable.

To define the loss function, we first select a set of ``target variables" $\{x\}$, such as the invariant masses of the intermediate states and invisible out-coming particles. For our specific example:

\vspace*{-15pt}
\begin{table}[h]
\begin{tabular}{ccl}
& \multicolumn{2}{c}{\quad\quad$\xleftrightarrow{\qquad \bigE \qquad} $ } \\
\multirow{4}{*}{$\bigB \left\{\mathclap{\begin{array}{c}\phantom{0}\\\phantom{0}\\\phantom{0}\\\phantom{0}\end{array}}\right.$} & \multicolumn{2}{c}{$x^{\# 1} = (m_{A_1},m_{B_1})^{\# 1} \oplus (m_{A_2},m_{B_2})^{\# 1}$} \\
& \multicolumn{2}{c}{\quad\quad\vdots} \\
& \multicolumn{2}{c}{$x^{\# N} = (m_{A_1},m_{B_1})^{\# N} \oplus (m_{A_2},m_{B_2})^{\# N}$}
\end{tabular}
\end{table}
Consider a batch of dataset consisting of $N$ training events. The Event-wise information, denoted as $\mathcal{E}$, is derived from the symmetric event topology. This implies that identical particle masses are consistent, making $\mathcal{E}$ an independent piece of information for each event.

On the other hand, the Batch-wise information, represented by $\mathcal{B}$, signifies that all training events are associated with the same physical event. We introduce auxiliary parameters, like $\tilde{m}_A$ and $ \tilde{m}_B$, to ensure that the masses across all events in a batch remain consistent (e.g., $m_{A_1}$ of all events are the same, and so on). This $\mathcal{B}$ information is dependent on the entire batch of events.
Finally, our loss function is defined as:
\begin{equation}
\label{eq:lossF}
\begin{aligned}
\mathcal{L}_{tot} &\equiv \frac{1}{|\mathcal{B}|}\sum_{i=1}^{|\mathcal{B}|} \left[ \sum_{f \in \{A,B\}} \mathcal{L}_{f}^{\#i}\right] \\
\mathcal{L}_{f}^{\#i} &\equiv  d_\mathcal{E}(m_{f_1}^{\#i}, m_{f_2}^{\#i})\\
&\qquad \qquad + \left[ d_\mathcal{B}(\widetilde{m}_{f}, m_{f_1}^{\#i}) + d_\mathcal{B}(\widetilde{m}_{f}, m_{f_2}^{\#i}) \right],
\end{aligned}
\end{equation}
where $|\mathcal{B}|$ represents the batch size, indicating the number of events in a batch $\mathcal{B}$, $\#i$ denotes the event index, and $f$ is the target variable, either $A$ or $B$. The functions $d_\mathcal{E}(x_1,x_2)$ and $d_\mathcal{B}(x_1,x_2)$ are distance functions for Event-wise and Batch-wise information, respectively. They satisfy mathematical conditions: (d1) $d(x_1,x_2)>0$ if $x_1\neq x_2$, $d(x_1,x_1)=0$, (d2) $d(x_1,x_2)=d(x_2,x_1)$, (d3) $d(x_1,x_2) \leq d(x_1,y) + d(y,x_2)$ for any $y$ in the sample. Various distance functions can be used, such as $d(x_1,x_2) = |x_1 -x_2|$, $|x_1-x_2|^2$, or $|x_1^2-x_2^2|$. The appropriate choice depends on the specific physical process under study.

\begin{figure}[t]
\includegraphics[width=0.48\textwidth]{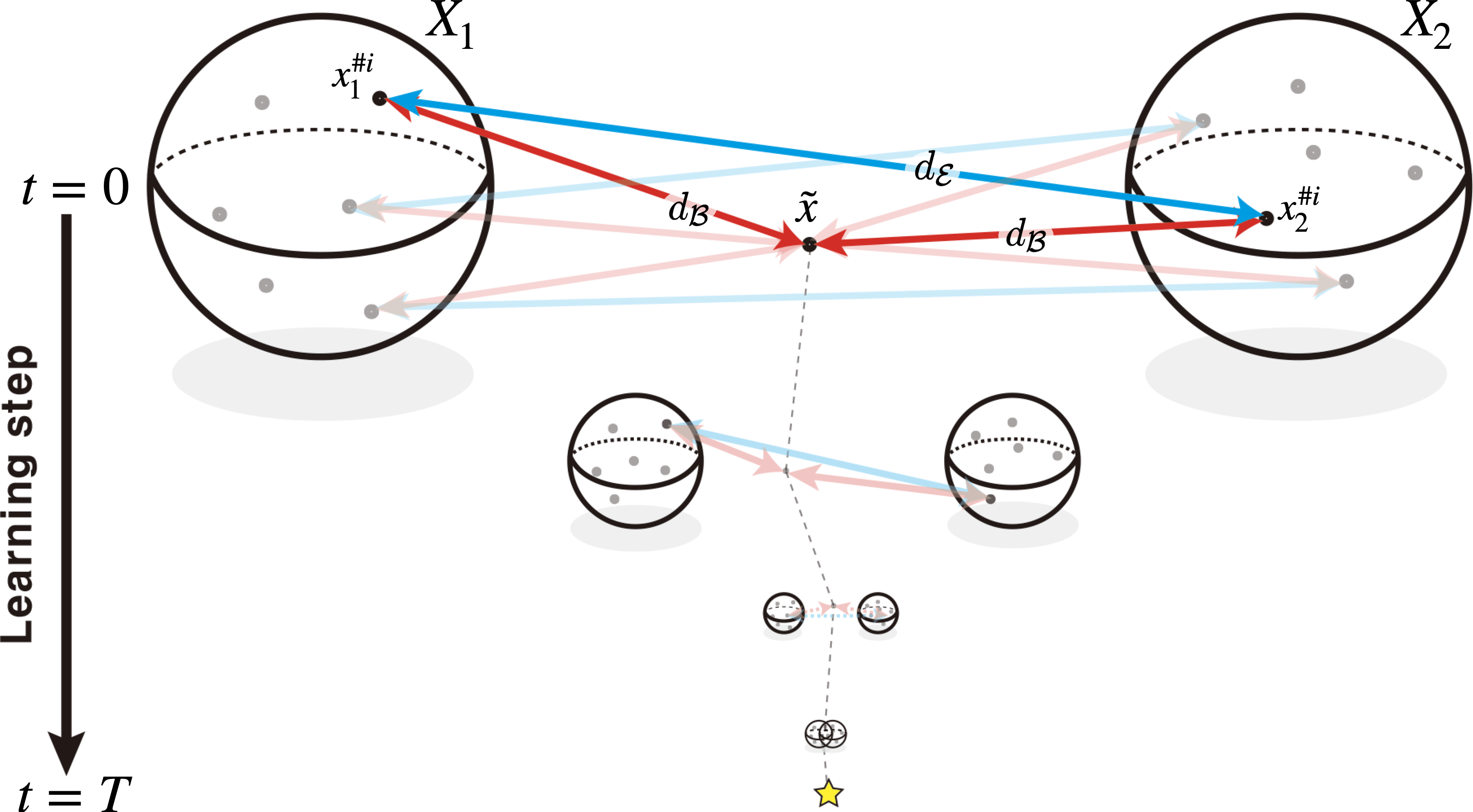}
 \caption{\label{train}\small Schematic representation of the role of the loss function in simultaneously bringing $d_{\mathcal{E}}$ (blue double-headed arrow) and $d_{\mathcal{B}}$ (red double-headed arrow) closer throughout the learning process $t=0$ to $t=T$.}
 \end{figure}

We illustrate the training procedure of \textsf{DeeLeMa} in FIG. \ref{train}. The target variable points ($x_1^{\#i}$, $x_2^{\#i}$) are represented within spaces $X_1$ and $X_2$, accompanied by the scalar value of the auxiliary parameter, $\tilde{x}$. By minimizing the loss function in Eq.(\ref{eq:lossF}) from the initial learning step at $t=0$ to the end of training at $t=T$, we ensure that spaces $X_1$ and $X_2$ come closer together. Additionally, the overall distribution of points within these spaces becomes more compact, leading to a reduction in their spread or dispersion. This compactness and reduction in dispersion are facilitated by the inclusion of the auxiliary parameter $\tilde{x}$.

For a comprehensive model implementation of \textsf{DeeLeMa}, refer to Appendix \ref{sec:appendix1}.

%%%%%%%%%%%%%%%%%%%%%%%%%%%%%
%%%%%%%%%%%%%%%%%%%%%%%%%%%%%

%%%%%%%%%%%%%%%%%%%%%%%%%%%%%%%%%%%%%%%%%%%%%%%%%%%%%%%%%%%%%%%%%%%%%%%%%%%%%%%%%%%%%%%%%%%%%%%%%%%%%%%%%%%%%%%%%%%%%%%%%%%%%%%%%%%%%%%%%%
\section{Test of \texorpdfstring{\textsf{D\MakeLowercase{ee}L\MakeLowercase{e}M\MakeLowercase{a}}}{DeeLeMa} performance}
\label{ex}
%====================

In pair production, practical experiments often encounter issues with the misidentification of branches. Termed the combinatorics problem, this complication can result in erroneous kinematic relations, leading to substantial uncertainties in the derived solutions. To quantify the extent of this contamination, we introduce the parameter $\mathcal{E}_C$, defined as the fraction of incorrectly assigned events relative to the overall number of events, expressed as:
\begin{align}
\mathcal{E}_C \equiv \frac{\text{wrong}}{\text{wrong}+\text{correct}}.
\end{align}

We assess the efficacy of \deelema through three distinctive test runs:
\begin{itemize}
\item Test run (A) is conducted using a toy model featuring fixed values of $m_A=1000~{\rm GeV}$, $m_B=800~{\rm GeV}$, and $m_C=700~{\rm GeV}$, with no combinatorial errors ($\mathcal{E}_C=0$).
\item Test run (B) mirrors (A) but incorporates varying rates of combinatorial errors, specifically $\mathcal{E}_C=0,~10\%,~20\%,~50\%$. This test aims to investigate the influence of combinatorial errors on the performance of \textsf{DeeLeMa}.
\item Test run (C) is executed on the standard model $t\bar{t}$ and $t \to W b \to (\ell \nu) b$ processes, encompassing $\mathcal{E}_C=20\%$ and accounting for detector smearing effects. We consider this test run to closely simulate a realistic scenario.
\end{itemize}
We compare the results with those obtained using other existing methods: the transverse mass variable $M_{T2}$ and the on-shell constrained invariant mass variables $M_{2CC}$, which use similar constraints as \textsf{DeeLeMa}. We use the \textsf{YAM2} package \cite{Park:2020bsu} to calculate $M_2$ optimally.

\subsection{Toy model }

\subsubsection{Toy model test with no contamination ($\mathcal{E}_C=0$)}

\begin{figure}[h]
	\includegraphics[width=0.46\textwidth]{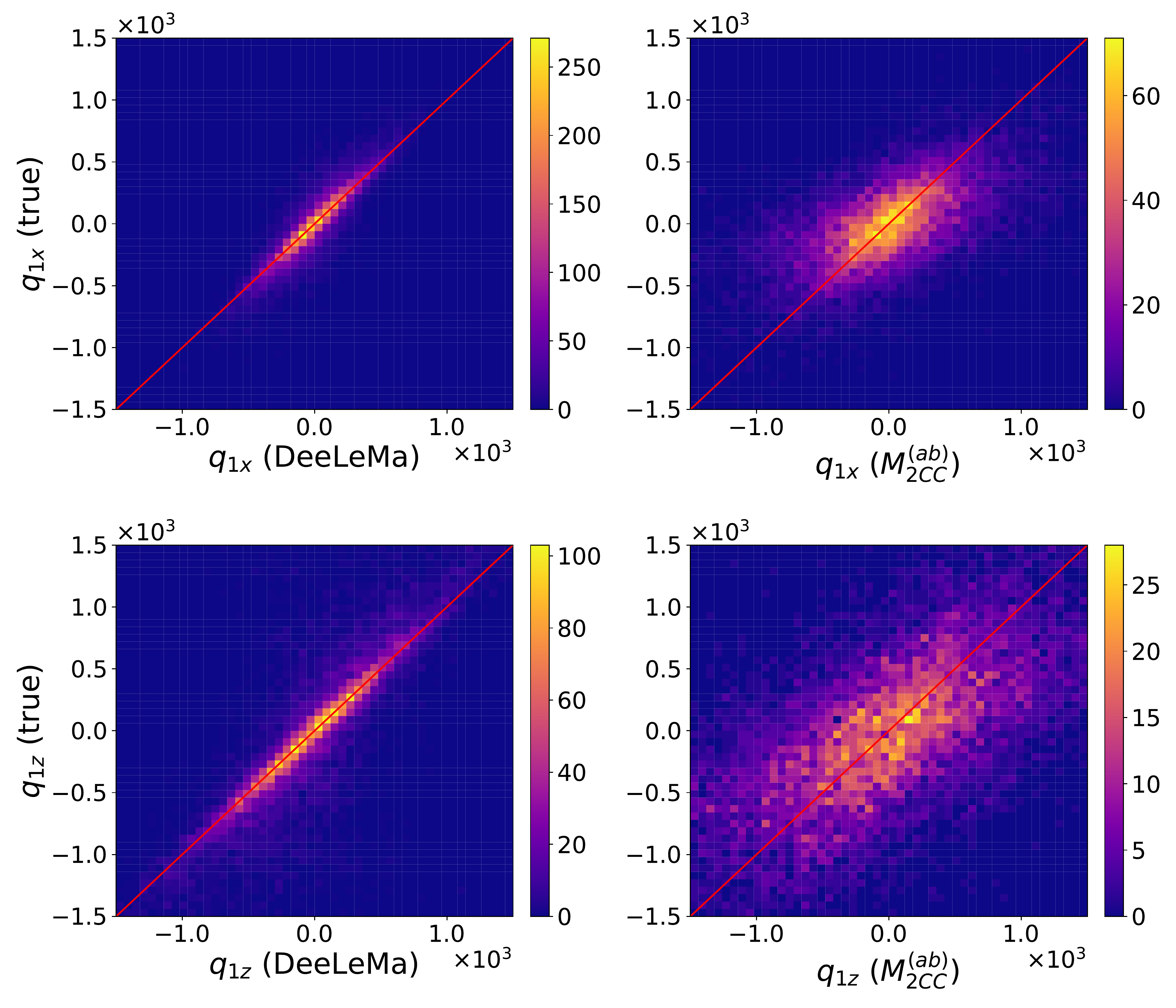}
	\caption{\label{fig:toy_mom}\small [Toy] The correlation heatmap of the reconstructed momenta and the true momenta from the \deelema (left) and $M_{2CC}^{(ab)}$ (right) for the toy example with $\mathcal{E}_C=0$.}
\end{figure}

We selected narrow width values for $m_A$, $m_B$, and $m_C$ at 1000 GeV, 800 GeV, and 700 GeV, respectively. % since the finite width can adversely affect performance.
 The correlation heatmap in FIG. \ref{fig:toy_mom} displays the relationship between the reconstructed momenta (horizontal axis) and the true momenta (vertical axis) for the \deelema method (left) and the $M_{2CC}^{(ab)}$ method (right) applied to the toy example with $\mathcal{E}_C=0$. Ideally, the diagonal line (red, solid line) should represent perfect efficiency with $p_{recon.}=p_{true}$. As shown in the figure, the \deelema method (left) exhibits a strong diagonal correlation pattern, indicating high accuracy in reconstructing the momenta. In contrast, the $M_{2CC}^{(ab)}$ method (right) shows a weaker and more scattered correlation pattern, implying a lower accuracy in momentum reconstruction. This demonstrates the superior performance of \deelema over traditional methods.

 \begin{figure}[h]
	\includegraphics[width=0.46\textwidth]{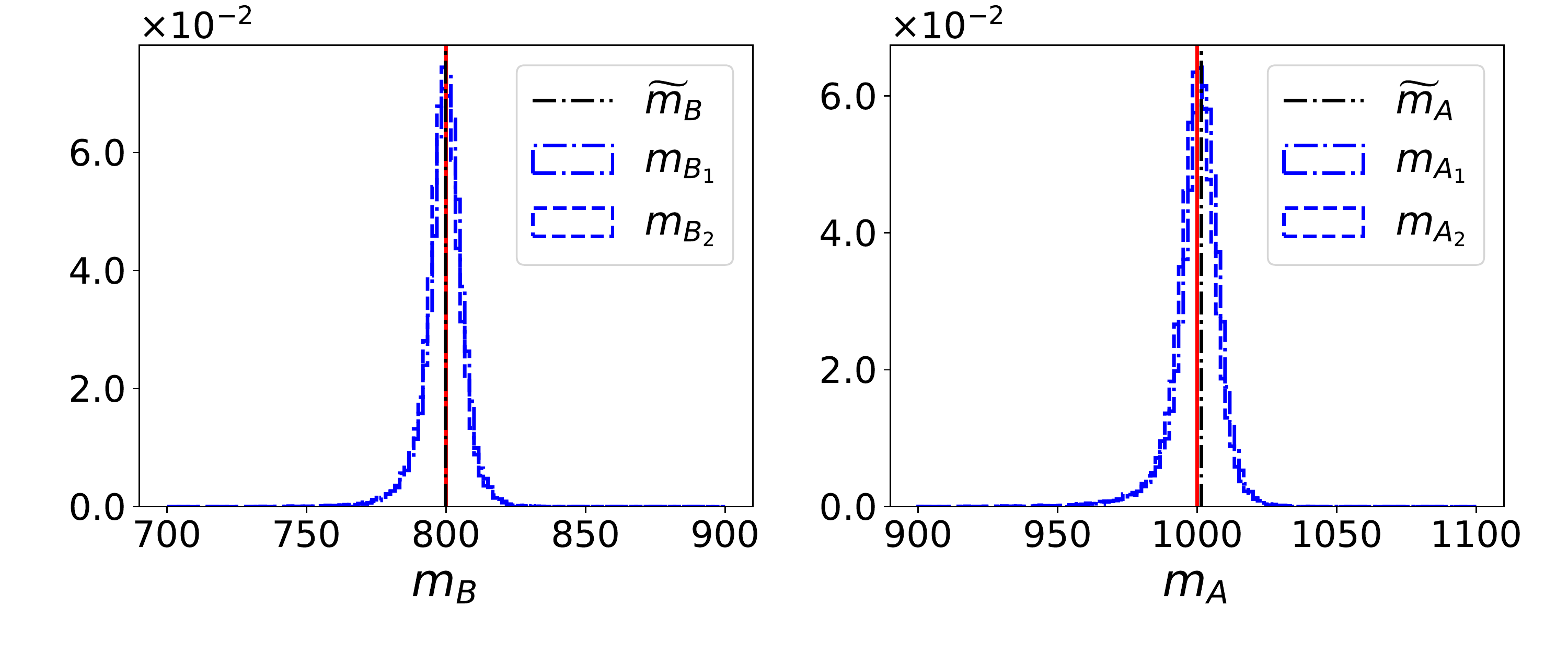}
	\includegraphics[width=0.46\textwidth]{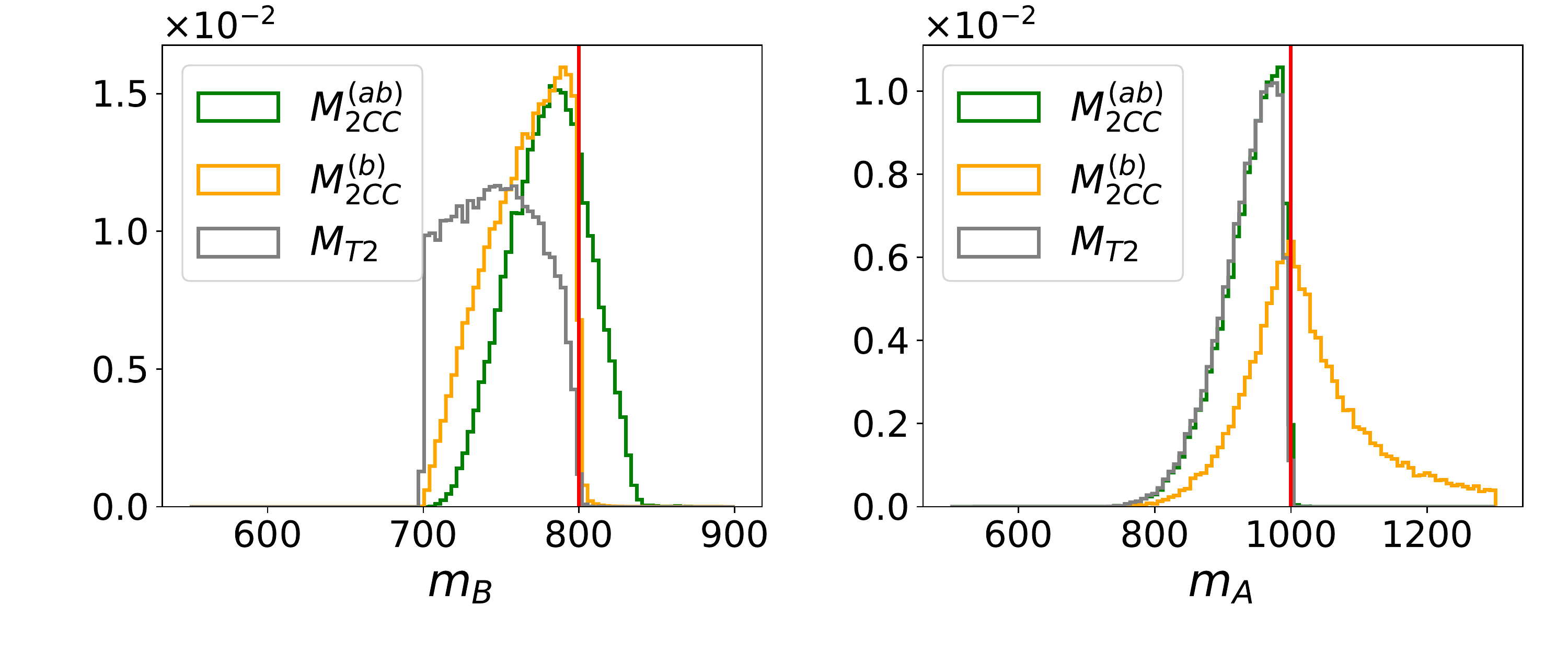}
	\caption{\label{fig:toy_m}\small [Toy] The reconstructed mass distributions of $B$ and $A$ using \deelema (upper), and $M_{T2}$, $M^{(b)}_{2CC}$ and $M^{(ab)}_{2CC}$ (bottom) for the toy example with $\mathcal{E}_C=0$.}
\end{figure}

%%%%%%%%%%%%%% Mass %%%%%%%%%%%%%% 

The upper panel of FIG. \ref{fig:toy_m} shows the reconstructed mass distributions of $B$ and $A$ obtained with \deelema for the toy example with $\mathcal{E}_C=0$. The blue dashed lines indicate the reconstructed masses of $m_{B_{1,2}}$ and $m_{A_{1,2}}$, respectively. The red vertical lines indicate the true masses, and the black dashed-dotted line shows the auxiliary mass after training. In the bottom panel, we compare the results with two existing methods based on $M_{T2}$ (gray) and $M_{2CC}$ with suitable subsystems $(b)$ and $(ab)$ (orange and green), respectively~\cite{Burns:2008va}. We can see that the reconstructed mass distributions with \deelema are well centered around the true values, while the $M_{T2}$ distribution shows the physical mass at the end-point of the distribution, which often causes errors. The $M_{2CC}^{(b)}$ for $m_B$ and the $M_{2CC}^{(ab)}$ for $m_A$ show slightly improved performances, but still \deelema provides the best results.

The disparity arises from the manner in which global information is assimilated during the machine learning training phase, primarily facilitated through the auxiliary parameter $\tilde{x}$. Conversely, in the context of the $M_{T2}$ or $M_{2}$ method, global information is solely derived from statistical outcomes, primarily centered around kinematic endpoints. While numerous events are typically clustered around these endpoints, leading to a reconstruction of momenta close to the actual values, there is a lack of subsequent optimization within the $M_{T2}$ or $M_{2}$ based reconstruction process.

Consequently, the precision is notably diminished, with the kinematic endpoints becoming less distinct, particularly when grappling with combinatorial ambiguities and accounting for the effects of detector smearing. Subsequently, this degradation in accuracy will be demonstrated in the subsequent sections.

\begin{table*}[t]
\setlength{\tabcolsep}{8pt}
\renewcommand{\arraystretch}{1.2}
\begin{tabular}{c||cc||cccc}
\hline\hline
$\mathcal{E}_C$  & \multicolumn{2}{c||}{$\widetilde{m}_X$ [~GeV~]} & \multicolumn{4}{c}{$m_X \pm \sigma$ [~GeV~] } \\ \hline
      [~\%~] & \multicolumn{1}{c|}{$\widetilde{m}_A$} & \multicolumn{1}{c||}{$\widetilde{m}_{B}$} & \multicolumn{1}{c|}{$m_{A_1}$} & \multicolumn{1}{c|}{$m_{A_2}$}  & \multicolumn{1}{c|}{$m_{B_1}$} & $m_{B_2}$ \\ \hline\hline
$0$          & \multicolumn{1}{c|}{1001.34}  & \multicolumn{1}{c||}{799.95}  & \multicolumn{1}{c|}{1000.45 $\pm$ 13.31}  & \multicolumn{1}{c|}{999.93 $\pm$ 13.59}  & \multicolumn{1}{c|}{799.59 $\pm$ 8.95}    &   799.42 $\pm$ 9.05  \\ \hline
$10$         & \multicolumn{1}{c|}{1001.46}  & \multicolumn{1}{c||}{800.47}  & \multicolumn{1}{c|}{1007.41 $\pm$ 32.21}  & \multicolumn{1}{c|}{1007.18 $\pm$ 31.90}  & \multicolumn{1}{c|}{802.26 $\pm$16.79}    &   802.11 $\pm$ 16.55  \\ \hline
$20$         & \multicolumn{1}{c|}{1005.16}  & \multicolumn{1}{c||}{802.25}  & \multicolumn{1}{c|}{1013.56 $\pm$ 43.24}  & \multicolumn{1}{c|}{1013.14 $\pm$ 42.24}  & \multicolumn{1}{c|}{804.59 $\pm$ 21.82}    &   804.41 $\pm$ 21.75  \\ \hline
$50$         & \multicolumn{1}{c|}{1010.73}  & \multicolumn{1}{c||}{807.61}  & \multicolumn{1}{c|}{1028.94 $\pm$ 62.56}  & \multicolumn{1}{c|}{1029.27 $\pm$ 61.39}  & \multicolumn{1}{c|}{810.87 $\pm$ 31.80}    &   810.97 $\pm$ 31.59  \\ \hline\hline
\end{tabular}
\caption{\small \label{table1} The summary table for combinatorial efficiency $\mathcal{E}_C$.}
\end{table*}
%

%%%%%%%%%%%%%% Combinatoric %%%%%%%%%%%%%% 
\subsubsection{Toy model test with contamination  ($\mathcal{E}_C>0$)}

To explicitly see the effect of combinatorics contamination, we conducted comprehensive test runs incorporating the possibility of combinatorial errors, with a concise summary of \deelema's performance presented in TABLE \ref{table1}. In these instances, the peak positions have displayed a slight shift towards larger values, owing to the influence of inaccurately assigned data implying a relatively higher mass. Despite accommodating up to $20\%$ in combinatorial errors, \deelema exhibits sustained resilience and commendable performance, accurately reconstructing masses within the $5-10\%$ range of the true values.

Notably, for cases where $\mathcal{E}_C\leq 20\%$, the reconstructed masses are within the vicinity of $\mathcal{O}(1)\%$ of the true values, attesting to \deelema's remarkable ability to mitigate the impacts of combinatorial challenges effectively. Collectively, our findings underscore \deelema's reliability and robustness as a method proficient in the precise reconstruction of masses, even in the face of demanding conditions prevalent in collider environments.

%%%%%%%%%%%%%%%%%%%%%%%%%%%%%%%%%%%%%%%%%%%%%%

\subsection{Realistic test with standard model $t\bar{t}$}

%==============
 \begin{figure}[h]
	\includegraphics[width=0.46\textwidth]{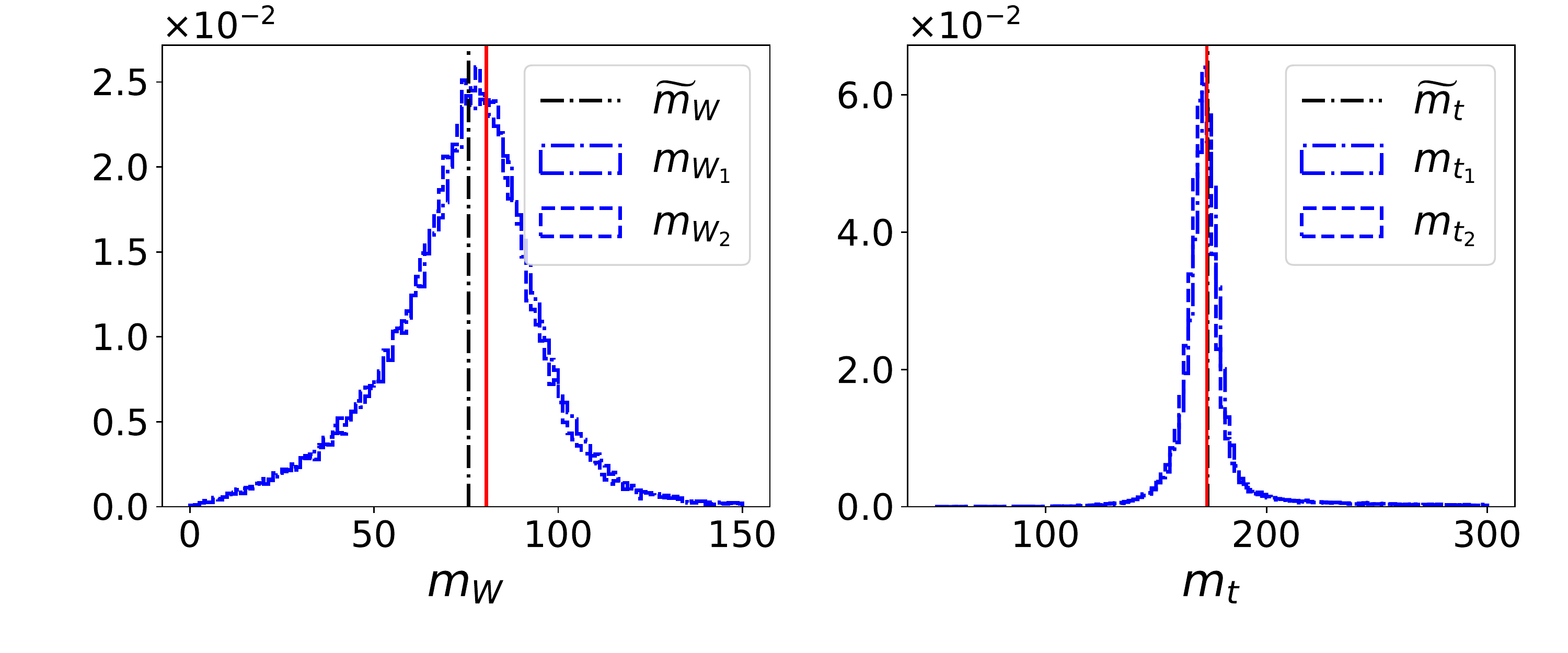}
	\includegraphics[width=0.46\textwidth]{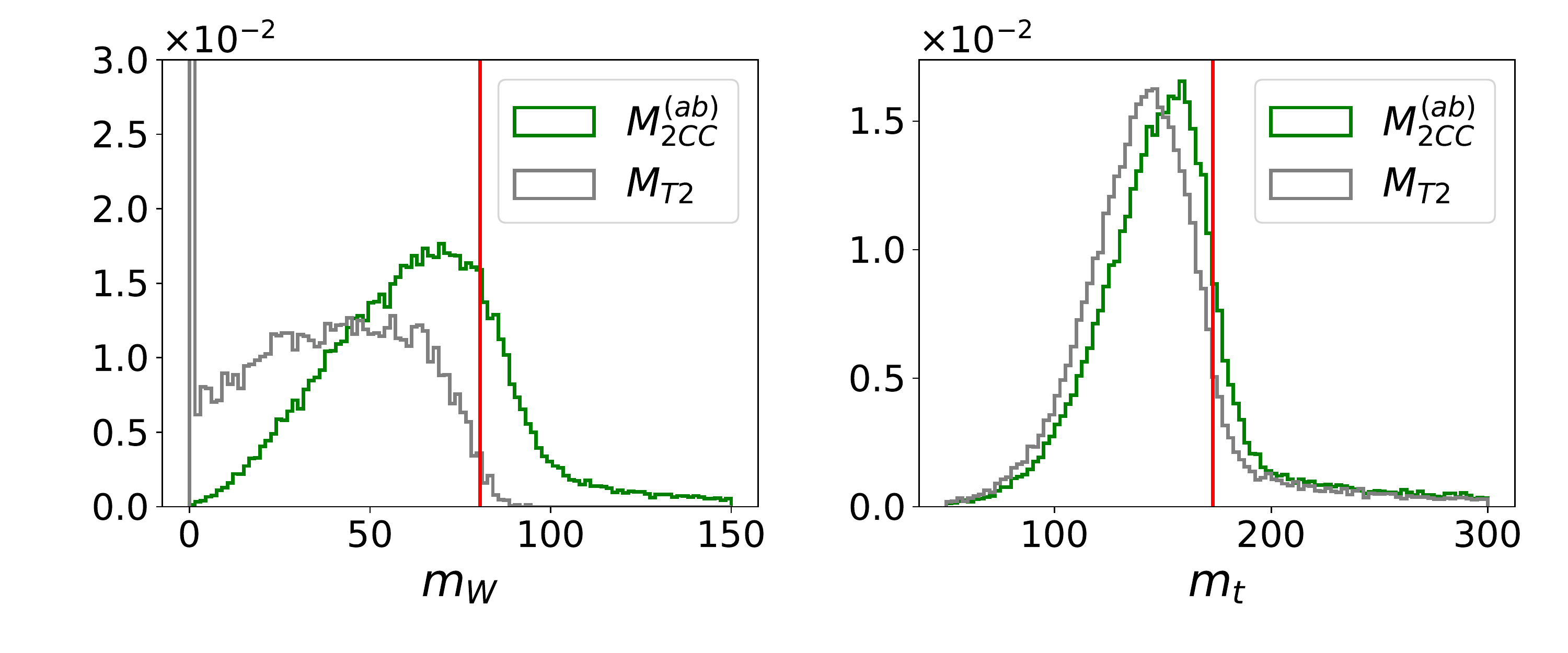}
	\caption{\label{fig:ttbar_m}\small [Realistic $t\bar{t}$] The reconstructed mass distributions of $B$ and $A$ using \deelema (upper), and $M_{T2}$, $M^{(b)}_{2CC}$ and $M^{(ab)}_{2CC}$ (bottom) for the $t\bar{t}$ example.}
\end{figure}
 %==============

We finally present the results of our investigation on a more realistic case, the top quark pair production at the LHC, where top quarks decay semi-leptonically as $t\bar{t} \rightarrow (b\ W^+)\ (\bar{b}\ W^-) \rightarrow (b\ \ell^+\ \nu)\ (\bar{b}\ \ell^-\ \bar{\nu})$. In this case, we consider finite width effects with $\sigma_t = 1.4915$ GeV, $\sigma_W=2.0476$ GeV, and $m_t = 173.0$ GeV, $m_W = 80.4190$ GeV for the top quark and $W$ boson, respectively. Moreover, we also account for the uncertainties related to the detector resolution.
We simulated detector effects by applying Gaussian smearing to the momenta. However, to achieve more accurate results, we encourage the use of a more realistic detector simulation. For the two $b$ jets, we applied Gaussian smearing with jet $p_T$ values of $\{10, 20, 30, 50, 100, 400, 1000\}$ GeV and energy resolutions of $\{40, 28, 19, 13, 10, 6, 5\}\%$, respectively \cite{Kim:2019prx, CMS:2016lmd}. We took the combinatorial ambiguity at $\mathcal{E}_C=20\%$ for our simulation.

We present the results obtained using \deelema in Fig.~\ref{fig:ttbar_m} (upper). The distributions for the reconstructed masses ($m_t$, $m_W$) show robust peaks near the true values (red vertical line), albeit slightly widened. To compare the performance of \deelema with conventional methods, we also present the results obtained using $M_{2CC}^{(ab)}$ and $M_{T2}$ variables (lower). \deelema provides more accurate results compared to conventional methods. In conventional methods, we need to read the endpoints in the lower distributions, which can be challenging in practice due to realistic effects from finite widths, detector smearing, and combinatorial mismatches.

%%%%%%%%%%%%%%%%%%%%%%%%%%%%%%%%%%%%%%%%%%%%%%
\section{Conclusion}
We introduce \deelema, a deep learning-based approach to analyze high-energy particle collisions with multiple invisible particles. 
\deelema can reconstruct the event's invisible momenta and masses, even when multiple invisible particles are involved. Focusing on a  challenging problem with $(N_{vis},N_{inv})=(4,2)$, we demonstrate the efficiency of \deelema:
compared to conventional methods that rely on kinematic variables such as $M_{T2}$ or $M_2$, \deelema delivers a significant improvement in accuracy. 
The reconstructed masses show sharp peaks in the distribution, and the results are robust against the combinatorial problem of misidentification of final state particles and detector-smearing effects.
In conclusion, \deelema has the potential to contribute to advances in the field as a new solid tool.

%%%%%%%%%%%%%%%%%%%%%%%%%%%%%%%%%%%%%%%%%%%%%%
\section*{Acknowledgments}
The work is supported by the National Research Foundation of Korea NRF-2021R1A4A20
01897(SCP), NRF-2019R1A2C1089334 (SCP), NRF-2021R1A6A3A1303942811 (KB). DWK is supported in part by KIAS Individual Grant. No. PG076202. We thank David Shih, Gregor Kasieczka, Doojin Kim, K. C. Kong, Chan Beom Park, Myeonghun Park, Seodong Shin, and Junji Hisano.

\appendix
\section{The detail of Model} \label{sec:appendix1}

The \deelema is constructed using the \textsf{PyTorch} package \cite{NEURIPS2019_bdbca288} and the \textsf{Lightning} library \cite{Falcon_PyTorch_Lightning_2019} as the front-end, with the \textsf{Adam} optimizer \cite{kingma2017adam} for training. The model is trained on GPUs with a specified batch size and number of epochs as summarized in  TABLE.~\ref{table:hyper_parameter}. Additionally, we employ the \textsf{GELU} (Gaussian Error Linear Unit) activation function \cite{hendrycks2023gaussian} with a $\tanh$ approximation and apply batch normalization. The detailed architecture and hyperparameters are available on the associated GitHub page\footref{github}. %and in TABLE.~\ref{table:hyper_parameter}.

\begin{table}[ht]
\centering
\label{table:hyper_parameter}
\renewcommand{\arraystretch}{1.5}
\resizebox{\columnwidth}{!}
{
\begin{tabular}{c|cccccccccc}
\noalign{\smallskip}\noalign{\smallskip}\hline\hline
Model & $N_\text{node}$ & $N_\text{layer}$ & $\eta$ & $|\mathcal{B}|$ & epoch & $m_C$ & $\Delta m_B^{\text{init}}$ & $\Delta m_A^{\text{init}}$ & $d$ \\
\hline
Toy& $256$ & $5$ & $10^{-2}$ & $2048$ & $100$ & $700$ & $0.3$ & $0.3$ & $L_1$\\

$t\bar{t}$& 256 & 5 & $5\times10^{-4}$ & $2048$ & $100$ & $0$ & $0.3$ & $0.3$ & $L_1$\\

\hline
\hline
\end{tabular}
}
\caption{Hyperparameters used for the result plots in \ref{ex}. $N_\text{node}$ denotes the number of nodes in each hidden layer, $N_\text{layer}$ represents the number of hidden layers, $\eta$ refers to the learning rate, $|\mathcal{B}|$ is the batch size, and epoch signifies the number of epochs.}
\label{table:hyper_parameter}
\end{table}

\newpage
\bibliography{ref}

\end{document}